\documentclass[11pt,twoside]{article}
\usepackage{./asp2014}

\aspSuppressVolSlug
\resetcounters

\begin{document}

\title{Emission from the Centrifugal Magnetospheres of Magnetic B-type Stars}
\author{Matt Shultz,$^{1,2,3}$ Gregg Wade,$^3$ Thomas Rivinius,$^1$ Richard Townsend,$^4$ and the MiMeS Collaboration
\affil{$^1$European Southern Observatory, Santiago, Chile; \email{mshultz@astro.queensu.ca}}
\affil{$^2$Queen's University, Kingston, Ontario, Canada}
\affil{$^3$Royal Military College, Kingston, Ontario, Canada}
\affil{$^4$University of Wisconsin, Madison, Wisconsin, USA}}

% This section is for ADS Processing.  There must be one line per author.
\paperauthor{Matt~Shultz}{mshultz@astro.queensu.ca}{}{Queen's University}{Dept. of Physics, Engineering Physics, and Astronomy}{Kingston}{ON}{K7L 3N6}{Canada}
\paperauthor{Gregg~Wade}{Gregg.Wade@rmc.ca}{}{Royal Military College}{Dept. of Physics}{Kingston}{Ontario}{K7K 7B4}{Canada}
\paperauthor{Thomas~Rivinius}{triviniu@eso.org}{}{European Southern Observatory}{Astronomy}{Vitacura}{Santiago}{19001}{Chile}
\paperauthor{Richard~Townsend}{townsend@astro.wisc.edu}{}{University of Wisconsin}{Dept. of Astronomy}{Wisconsin}{Madison}{53706}{USA}

\begin{abstract}
Approximately 10\% of B-type stars possess strong magnetic fields, and of these, 25\% host centrifugal magnetospheres in which the radiative wind, magnetic field, and rotational support interact to form a dense circumstellar plasma visible in a variety of diagnostic lines. In this article we review the basic theory behind CMs, outline current theoretical and observational problems, compare the observational properties of CM host stars to those of classical Be stars, and finally present preliminary results of a population study aimed at clarifying the characteristics of this growing sub-class. 
\end{abstract}

\section{Introduction}

As noted in a recent review by \cite{2013AARv..21...69R}, there are numerous varieties of B-type stars displaying emission originating in their cirumstellar environments. First, and most common, are the classical Be stars, the primary subject of these proceedings, in which the emission originates in decretion disks. Other varieties include the B-type supergiants (where the emission originates in winds), Herbig Ae/Be stars (accretion disks), B[e] stars (dusty winds), and finally, the magnetic B-type stars (magnetospheres). It is this latter class that is the subject of this article. 

A principal result of the Magnetism in Massive Stars (MiMeS) Survey Component (SC) is that approximately 10\% of OBA stars possess strong ($\sim$1--10 kG) magnetic fields \citep{2012AIPC.1429...67G}. It should be noted that these statistics differ for different sub-classes of stars. For instance, HAe/Be stars show approximately the same incidence fraction as the general population \citep{2013MNRAS.429.1001A}. On the other hand, there is not a single instance of a magnetic Be star (see e.g. Wade et al., these proceedings). Similarly, magnetic fields have not been detected in BA supergiant stars \citep{shultz2014}. Finally, amongst the Bp stars, especially the He-weak and He-strong stars, the incidence fraction of magnetic fields is essentially 100\%. 

Magnetism amongst the hot stars is quite distinct from that seen in cool stars. First, magnetic fields are essentially ubiquitous amongst low-mass stars. Second, cool stars exhibit magnetic activity cycles similar to those of the Sun, as expected for dynamo-generated magnetic fields \citep{2009ARAA..47..333D}. In contrast, the magnetic fields of massive stars are extremely stable, exhibiting no variation between rotational cycles. This has led to their characterization as so-called `fossil' fields, left-overs of magnetic flux preserved from star formation regions, generated during the star formation process itself, or perhaps a consequence of binary mergers. 

The strong magnetic fields of the Bp stars have a number of consequences for photospheric and circumstellar structures that lead to numerous indirect diagnostics of stellar magnetism. All such diagnostics are, like the magnetic variations, synchronized to the stellar rotational period. The most ubiquitious are surface abundance spots, with over- and under-abundant patches associated with the magnetic poles or equators. The remaining diagnostics largely originate in magnetospheres: emission in optical H lines and ultraviolet lines; hard, strong X-ray emission; and radio emission. 

\section{Magnetic Wind Confinement}

All B-type stars have line-driven winds; however, on the main sequence, these winds are typically too weak for their emission to become detectable. However, under the right circumstances, wind plasma can interact with stellar magnetic fields in such a way as to become optically thick. 

The first requirement is that the magnetic field be strong enough to confine the wind. This is expressed by the magnetic wind confinement parameter $\eta_*$, which is simply the ratio of the magnetic energy density to the wind kinetic energy density \citep{ud2002}: 

\begin{equation}\label{etastar}
\eta_{*} \equiv \frac{B_{\rm eq}^{2}R_{*}^{2}}{\dot{M}v_{\infty}},
\end{equation}

\noindent where $B_{\rm eq}$ is the strength of the magnetic field evaluated at the stellar equator, $R_*$ is the stellar radius, $\dot{M}$ is the mass-loss rate, and $v_{\infty}$ is the wind terminal velocity. If $\eta_* > 1$, the wind is said to be magnetically confined. To get a sense of the {\em physical} extent of magnetic confinement, equation \ref{etastar} can be used to calculate the the Alfv\'en radius, that is, the maximum extent of closed magnetic loops \citep{ud2002}: 

\begin{equation}\label{ralf}
\frac{R_{\rm A}}{R_*} \approx 0.3 + (\eta_* + 0.25)^{1/4}.
\end{equation}

For a stellar magnetosphere to become visible, it is not sufficient that $\eta_* > 1$. In the absence of rotation, plasma confined within the magnetosphere will simply collide at the magnetic equator, stall, and fall back to the stellar surface under the influence of gravity. Since the plasma leaves the magnetosphere on dynamical timescales, the density will not rise to the point of optical thickness unless $\dot{M}$ is very high. Thus, it is only in the magnetic O-type stars that such `dynamical magnetospheres' (DMs) become visible (\citealt{petit2013}; see also Fig. \ref{ipod}). 

Rotation changes this situation. Since the magnetically confined plasma is forced to corotate with the photosphere out to $R_{\rm A}$, a centrifugal force acts on the plasma. The Kepler radius  $R_{\rm K}$ is defined as the radius at which gravitational centrifugal forces balance  \citep{town2005, ud2008}

\begin{equation}\label{rkep}
R_{\rm K} \equiv \left(\frac{GM}{\omega^2}\right)^{1/3},
\end{equation}

\noindent \noindent where $G$ is the gravitational constant, $M$ is the stellar mass, and $\omega$ is the angular rotational velocity, which can be determined from $R_*$ and $v\sin{i}$~if the inclination from the line of sight $i$ is known. If $R_{\rm K} < R_{\rm A}$, there is a region within which rotational support prevents the plasma from falling back to the star, while magnetic confinement prevents the plasma from leaving the magnetosphere. This is referred to as a `centrifugal magnetospheres' (CM; \citealt{petit2013}). 

\articlefiguretwo{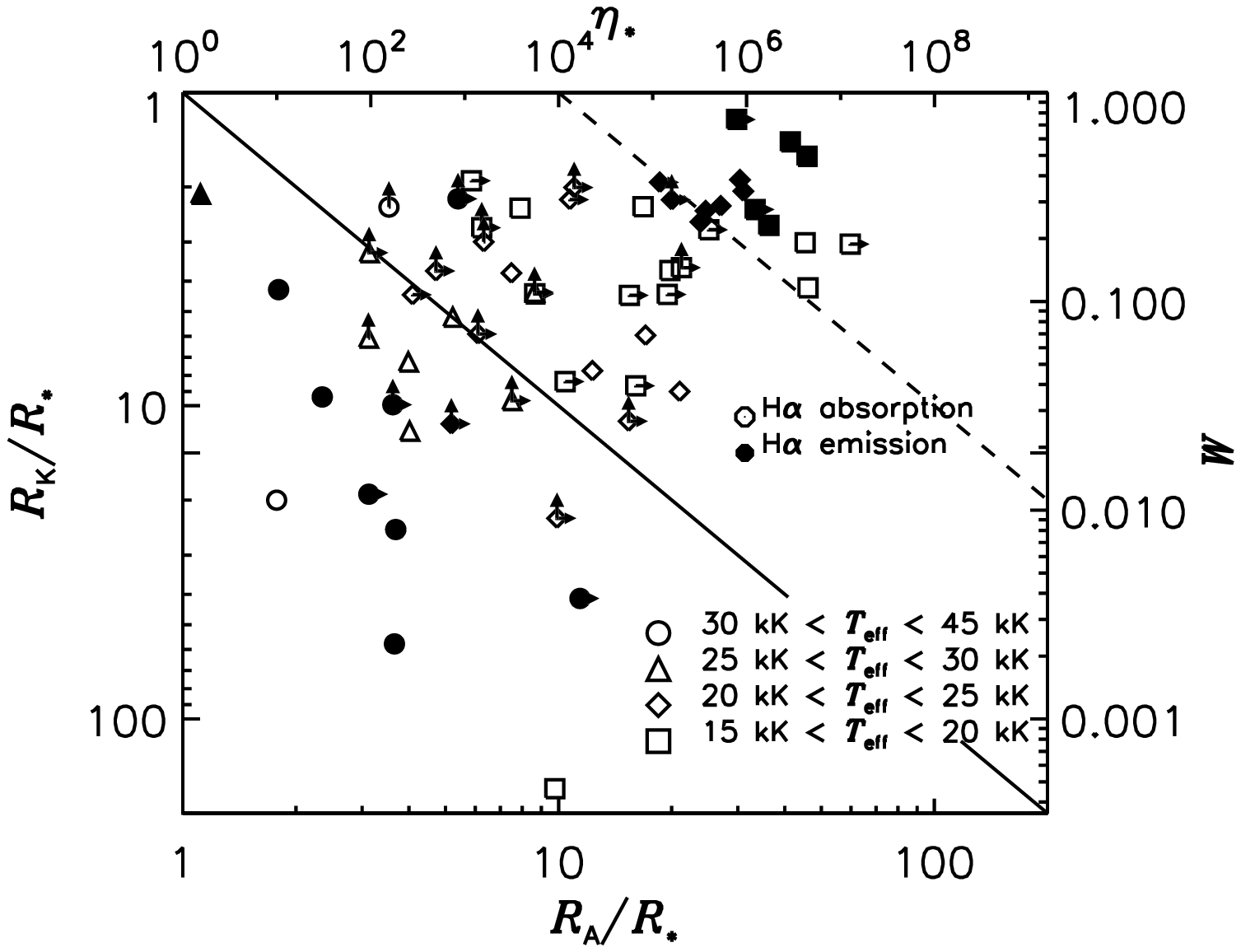}{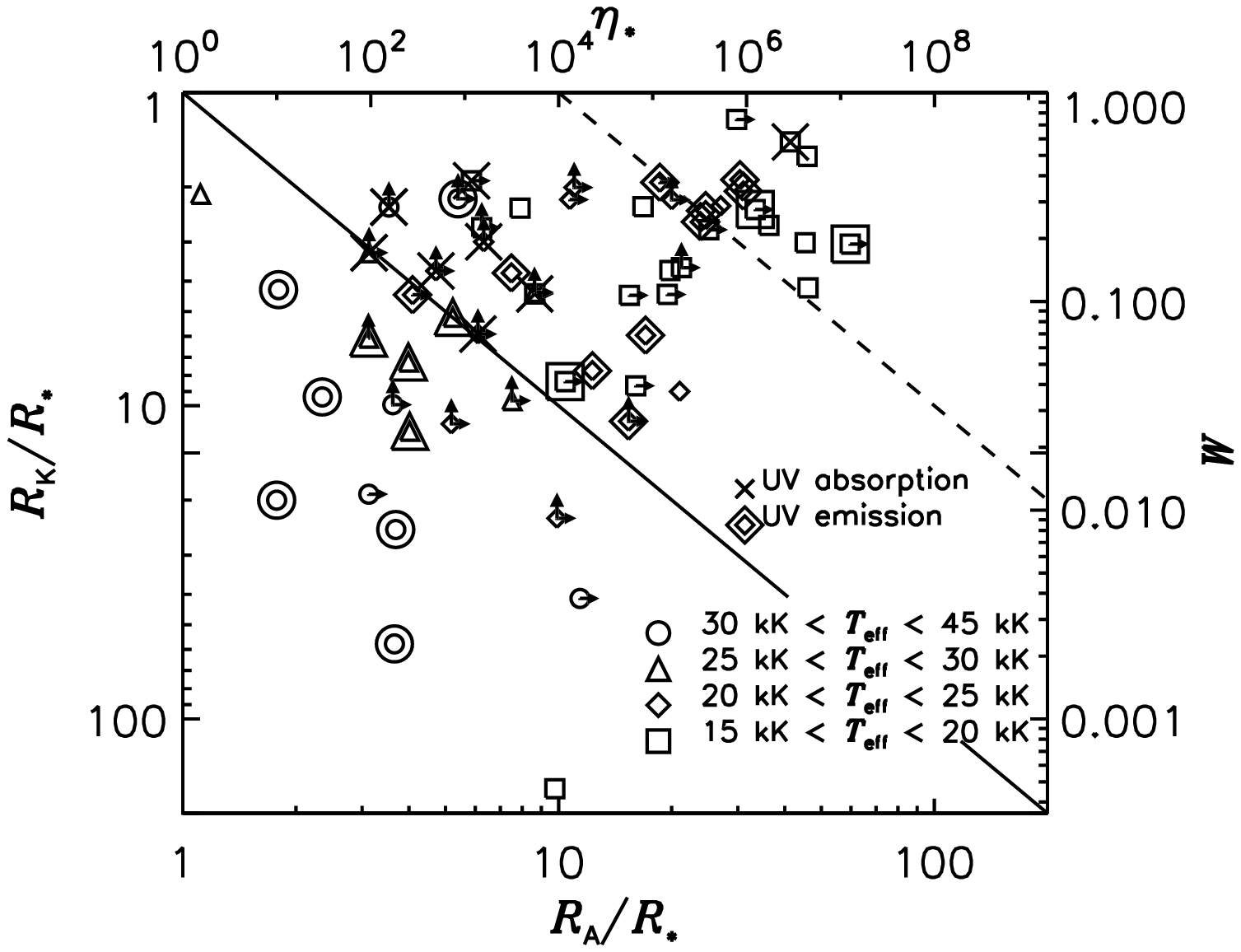}{ipod}{The rotation-confinement diagram. The solid diagonal line indicates $R_{\rm A} = R_{\rm K}$: stars below this line possess only DMs, stars above also possess CMs. The dashed diagonal line indicates $R_{\rm A} = 10R_{\rm K}$. {\em Left}: filled symbols indicate H$\alpha$ emission, open simples H$\alpha$ absorption. {\em Right}: Double symbols indicate UV emission, crosses UV absorption, single symbols stars for which UV data is not available.}

The dual importance of rotation and magnetic confinement is illustrated in the rotation-confinement diagram (\citealt{petit2013}, see also Fig. \ref{ipod}). This diagram arranges magnetic stars according to $R_{\rm A}$ and $R_{\rm K}$ (or equivalently, $\eta_*$ and the rotation parameter, $W$). As is apparent from Fig. \ref{ipod}, the magnetic B-type stars span a wide range of magnetospheric parameters, with $1 < R_{\rm A} < 60$, and $1 < R_{\rm K} < 100$. However, stars showing H$\alpha$ emission all have $R_{\rm A} >> R_{\rm K}$.

The conditions of extreme magnetic field strength and rapid rotation characterizing CM host stars are beyond the reach of numerical, time-dependent magnetohydrodynamics (MHD) simulations, as the high Alfv\'en speeds lead to short time steps that make the problem computationally intractable. However, CMs can be successfully modeled via the Rigidly Rotating Magnetosphere model (RRM; \citealt{town2005}), a time-independent, semi-analytic formalism which is able to treat arbitrary magnetic topologies. RRM computes an accumulation surface defined by the intersection of the magnetic field with the local minima of the gravitocentrifugal potential. Hydrostatic equlibrium is then assumed along each magnetic field line, with the plasma settling on the accumulation surface. For the canonical case of an oblique rotator (i.e., a dipolar magnetic field with an angle between the magnetic axis and the rotational axis), RRM predicts a warped disk morphology, with the densest regions corresponding to the intersections of the magnetic and rotational equators.  

While RRM does not perform radiative transfer, it is able to predict the approximate shape of emission lines, as well as the photometric depths of eclipses by the plasma clouds \citep{2005ApJ...630L..81T, 2008MNRAS.389..559T}. However, it makes no predictions about energy output in different electromagnetic regions. By relaxing the assumption of hydrostatic equilibrium, Rigid-Field Hydrodynamics (RFHD) is able to predict emission at UV, EUV, and X-ray energy levels (\citealt{2007MNRAS.382..139T}; see also Bard \& Townsend, these proceedings). 

\subsection{Plasma Leakage}

Obviously, plasma cannot continue to accumulate in the CM forever. However, the mechanism by which plasma escapes the CM remains the subject of debate. \cite{1984AA...138..421H} were the first to take up this question, considering thermal or dynamical equilibrium (which operates in DMs, but by definition not in CMs), diffusive mechanisms, and finally stochastic, violent ejection of the plasma. 

The simplest diffusive mechanism is ambipolar diffusion. This has been ruled out by the high ionization fraction and short mean free paths of neutral particles \citep{1984AA...138..421H, town2005}. 

The leading theoretical candidate for a leakage mechanism is a stochastic process referred to as `centrifugal breakout' (CB; \citealt{ud2006}). This occurs when the plasma density exceeds the magnetic field's ability to confine it. Under ideal MHD conditions, in particular `frozen flux' in which plasma and magnetic field lines move together, the magnetic field lines will now move with the plasma rather than vice-versa. An analytical estimate of CB timescales $t_{\rm B}$  was performed by \cite{town2005}, who found that for the prototypical CM host star $\sigma$ Ori E, $t_{\rm B} \sim 220$ yr. CB is also a feature of MHD simulations incorporating rotation \citep{ud2008}. 

Despite strong theoretical grounds for expecting CB, observational evidence is lacking. \cite{2013ApJ...769...33T} examined three weeks of MOST space photometry of $\sigma$ Ori E, finding no variation between rotational cycles. Additionally, broadband polarization observations of the same star by \cite{2013ApJ...766L...9C} indicated that the circumstellar plasma is concentrated more strongly at the intersections of the rotational and magnetic equators than expected from RRM. Whether this discrepancy is a consequence of plasma leakage, or of an overly simplified model of the photospheric magnetic field \citep{oks2012}, is a subject of active investigation \citep{2014arXiv1408.0627O}.

\articlefigure{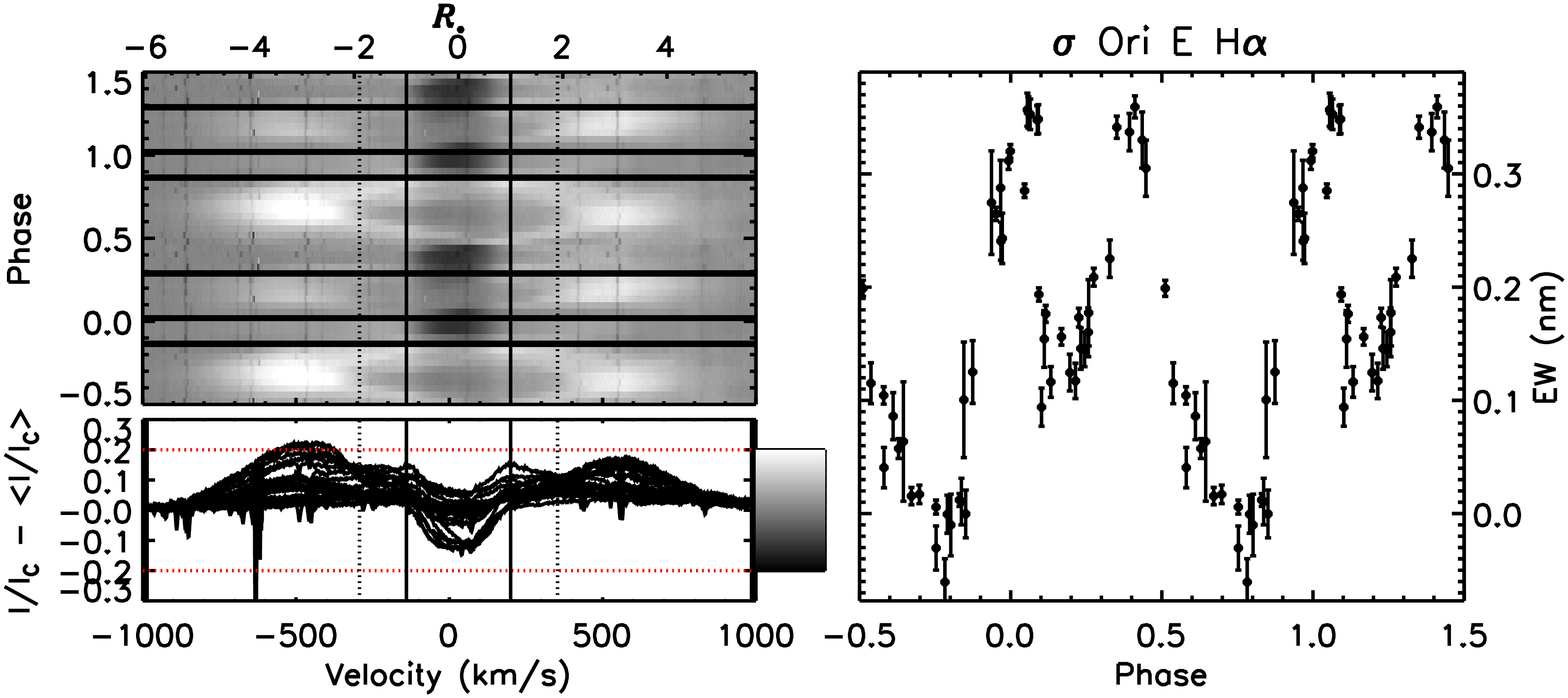}{dyn_ew}{{\em Left}: Dynamic spectrum of $\sigma$ Ori E's H$\alpha$ line. The bottom panel shows the 1D residual flux as compared to a model photospheric spectrum (shown in Fig. \ref{halpha_mosaic}), with the mapping to the grayscale colour bar to the right. The top panel shows the residual flux as a function of rotational phase. Abssicae are in units of velocity (bottom) and stellar radii (top). Vertical solid lines indicate $\pm v\sin{i}$, dashed lines $\pm R_{\rm K}$. {\em Right}: EW as a function of a phase.}

\articlefigure{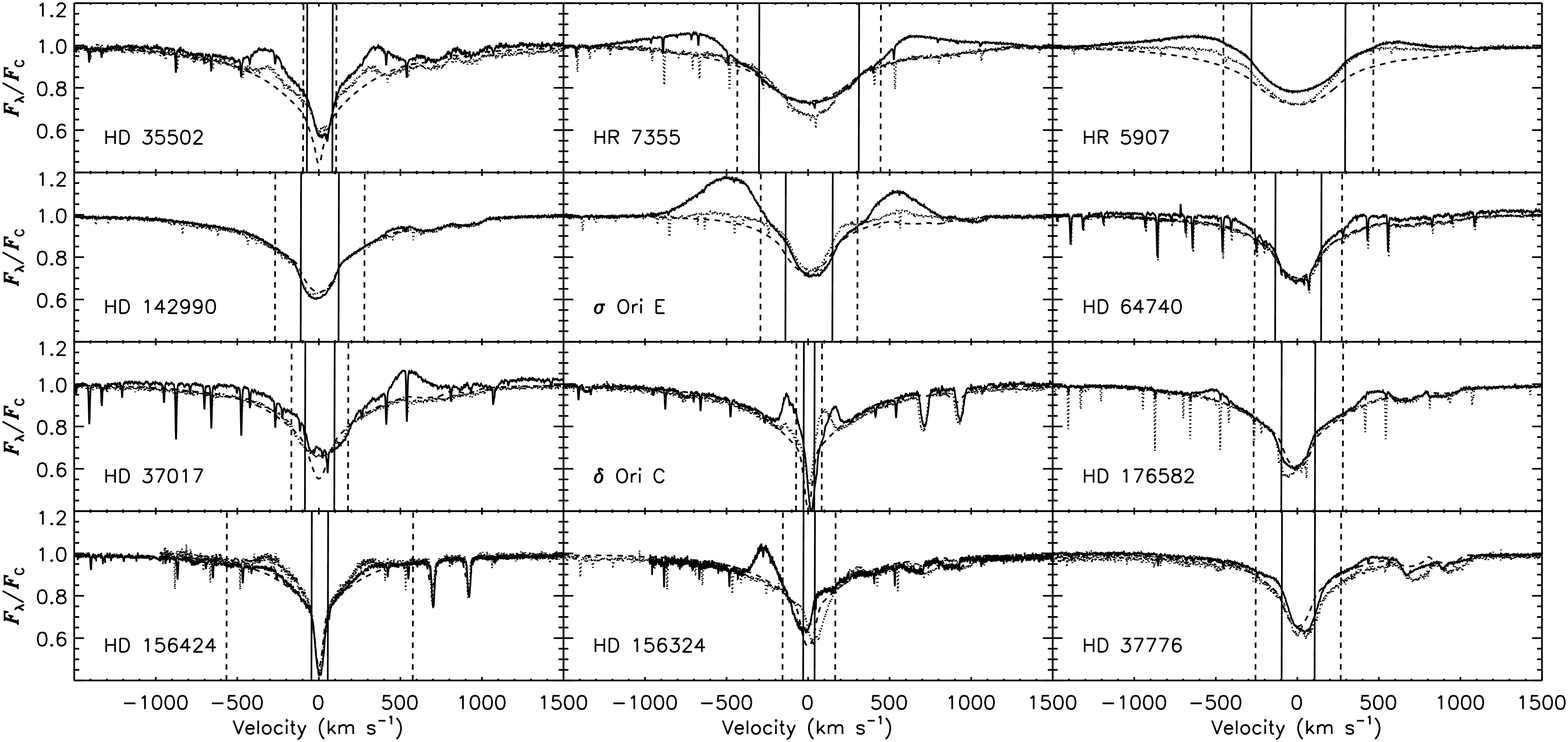}{halpha_mosaic}{Mosaic of H$\alpha$ lines for various CM host stars. Each panel shows H$\alpha$ at maximum emission and maximum absorption. Curved dashes lines indicate synthetic photospheric spectra. Vertical solid lines indicate $\pm v\sin{i}$, dashed lines $\pm R_{\rm K}$. Note that in the cases of HD 156324 and HD 156424, only upper limits are available for $R_{\rm K}$.}

\section{Variability}

Fig. \ref{dyn_ew} shows a dynamic spectrum and equivalent width (EW) measurements of the H$\alpha$ line of $\sigma$ Ori E, phased with the spindown ephemeris presented by \cite{2010ApJ...714L.318T}. Clearly visible are the high-velocity emission at quadrature phases (when the plasma clouds are projected on either side of the star), and the absorption excesses in the rotationally broadened core of the line (when the plasma clouds occult the stellar disk). As might be expected, the occultations also lead to photometric variability. 

Similar patterns have been observed in numerous other CM host stars: $\delta$ Ori C \citep{leone2010}, HR 7355 \citep{2013MNRAS.429..177R}, HR 5907 \citep{grun2012}, HD 176582 \citep{bohl2011}, and HD 37017 \citep{1993AA...273..509L} have all been studied at some level of detail. These, and other examples are shown in Fig. \ref{halpha_mosaic}. The basic pattern, of emission peaking at high velocites at quadrature phases, is the same in all cases. Intriguingly, the emission often peaks at around $1.3R_{\rm K}$: this phenomenon is also a feature of RFHD simulations, as discussed in more detail by Bard \& Townsend (these proceedings). 

As is apparent from Fig. \ref{ipod}, many stars that do not show H$\alpha$ emission do emit at UV wavelengths. While UV data is somewhat scarce, of those for which IUE observations are available, 19/27 (or 70\%) show emission, as compared to 14/52 (26\%) which show H$\alpha$ emission. 

\section{Comparison with classical Be stars}

Superficially, classical Be stars and CM host stars may seem quite similar: both are typically rapid rotators, and both display emission in H Balmer lines and wind-sensitive UV lines. However, on closer examination, these stars are easily distingushed. 

First, emission from CMs is quite weak: the strongest emission is seen in $\sigma$ Ori E, at about 20\% of the continuum (see Fig. \ref{halpha_mosaic}), whereas in Be stars emission at $10\times$ the level of the continuum is not uncommon. Similarly, the photometric variation due to CMs is generally less than 0.05 mag, while for Be stars it is much stronger, around 0.5 mag. 

Second, the rigid corotation of CMs means that the strongest emission is found at typically $2-3\times v\sin{i}$. Conversely, in classical Be stars the emission originates in a viscous decretion disk in Keplerian orbit, and is consequently confined within $\pm v\sin{i}$, peaking at or near zero velocity. 

Third, the timescales of variability are quite distinct. With the exception of pulsating stars, all variation around magnetic B-type stars is due to rotational modulation: variable emission from CMs is therefore strictly periodic, and quite rapid (since only stars with $P_{\rm rot} < 1.5$ d possess detectable CMs). While there is some degree of rotational modulation of Be star emission, the strongest variability is due to intrinsic evolution (buildup and dissipation) of the decretion disk, a process that can take months or years. This variability is moreover often aperiodic or at best semi-periodic, in strong contrast to that of CM host stars. 

\articlefigure{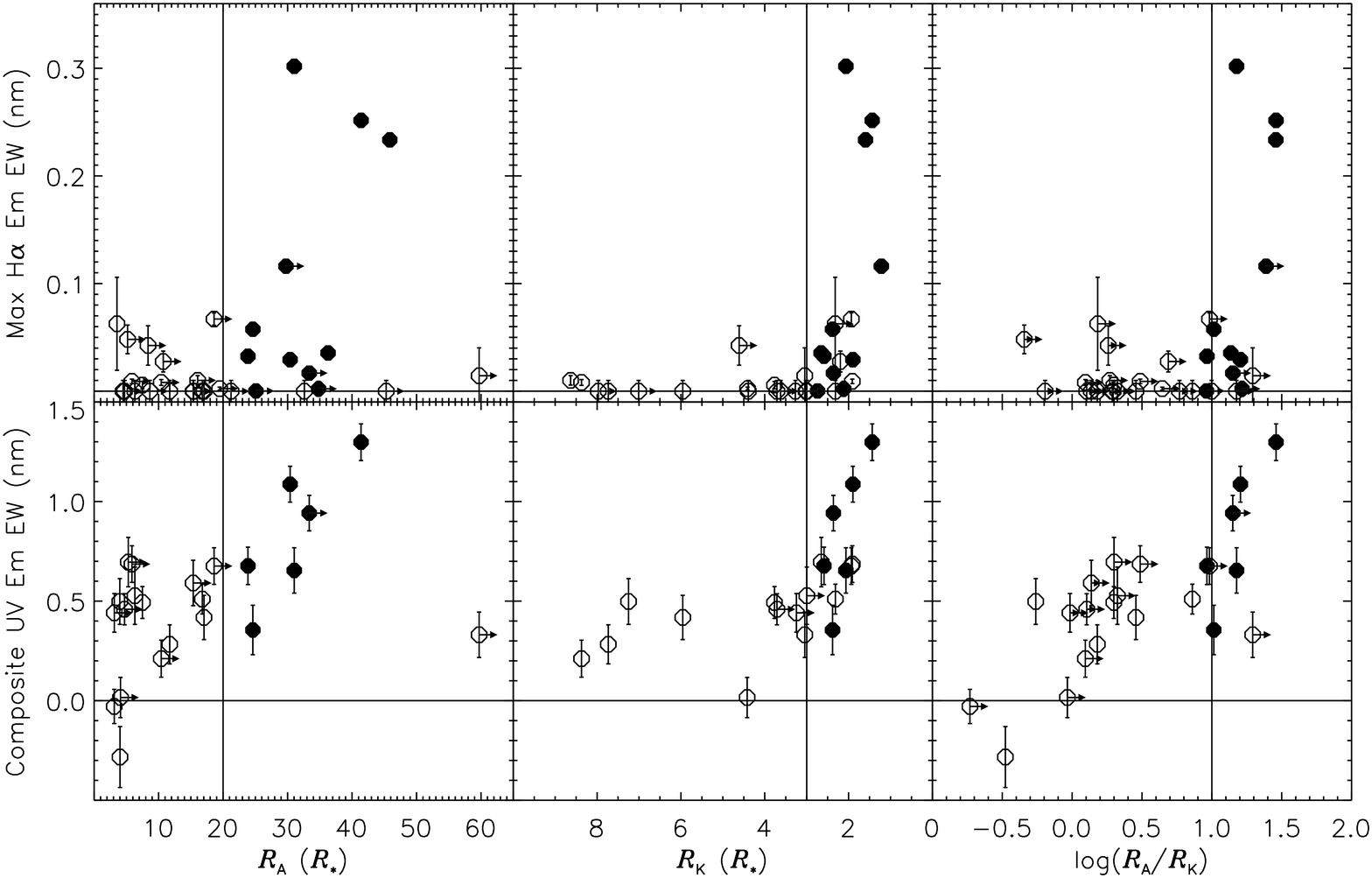}{halpha_uv_maxem}{Maximum emission EW for H$\alpha$ (top) and composite UV (bottom), as functions of (left--right) $R_{\rm A}$, $R_{\rm K}$, and $\log{(R_{\rm A}/R_{\rm K}})$. Vertical lines indicate the thresholds discussed in the text. Filled symbols correspond to stars with $R_{\rm A}$ ($R_{\rm K}$) above (below) the threshold values. Arrows indicate stars for which only limiting values exist.}

\section{The Population Study}

While RRM has been qualitatively successful, there are numerous open questions. Not the least of these is plasma leakage, as discussed above. A potentially connected mystery is why emission should only be seen in the stars with the most extreme rotational and magnetic conditions. 

In order to explore these isses, we are conducting a population study of all known early B-type magnetic stars, both those listed by \cite{petit2013}, and those that have been discovered since. The sample currently includes 45 stars (with the magnetic Herbig Be stars having been left out). The primary observational data includes 1198 high-resolution spectropolarimetric sequences acquired with ESPaDOnS, Narval, and HARPSpol, supplemented by optical FEROS (152 spectra) and UVES (418 spectra) data, and archival IUE spectroscopy (509 spectra). 

The majority of the spectropolarimetry was acquired by the MiMeS Large Programs; however, it also includes 185 ESPaDOnS  sequences acquired in PI programs aimed at observing stars neglected by the MiMeS Targeted Component. These data have been used to determine rotational periods for 8 additional stars. They have also led to the discovery of a new CM star, ALS 3694 (Shultz et al., these proceedings). 

The first aim of this project is to clarify the stellar, atmospheric, rotational, and magnetic properties of the sample stars, so as to more accurately place them on the rotation confinement diagram. The second goal is to examine the emission properties of the population in greater detail, and thus to conduct a systematic confrontation with RRM predictions. Here we consider the progress made towards the second goal, as being more germane to the theme of this article. 

We first measured the emission EWs of the sample stars, by subtracting synthetic non-LTE TLUSTY spectra \citep{2007ApJS..169...83L} from the observed spectra, as illustrated in Fig. \ref{halpha_mosaic}. Line cores were masked out, in order to avoid distortion of the results by occultation effects. Fig. \ref{halpha_uv_maxem} shows the maximum H$\alpha$ EWs regressed against $R_{\rm A}$, $R_{\rm K}$, and $\log{(R_{\rm A}/R_{\rm K}})$. There is no clear trend of increasing emission with any of these parameters; however, there do appear to be clear {\em thresholds} for emission, with no stars showing emission if $R_{\rm A} < 20$, $R_{\rm K} > 3$, or $\log{(R_{\rm A}/R_{\rm K}}) < 1$. 

Using IUE spectra, a similar operation was performed on the wind-sensitive doublets N {\sc v} 1240 nm, Si {\sc iv} 1400 nm, C {\sc iv} 1549 nm, and Al {\sc iii} 1858 nm. The bottom panels of Fig. \ref{halpha_uv_maxem} show the composite emission EWs. Emission is much stronger in the UV lines than in H$\alpha$, and there is moreover more evidence of a trend with increasing $R_{\rm A}$ and decreasing $R_{\rm K}$. 

We performed similar measurements on H$\beta$ and H$\gamma$, which we then used to measure the circumstellar plasma density using Balmer decrements, for the 6 stars for which the emission is strong enough to do so: $\delta$ Ori C, HR 7355, HD 37017, $\sigma$ Ori E, HD 156424, and HD 176582. The volume density is identical within error bars to a mean value of $\log{N} = 12.6 \pm 0.1$. This is consistent with previous results for $\delta$ Ori C \citep{leone2010} and HR 7355 \citep{2013MNRAS.429..177R}. 

These results are incompatible with a CB scenario. First, CB predicts a clear trend of increasing plasma density with increasing $R_{\rm A}$ (as stronger magnetic fields are able to confine more plasma) and decreasing $R_{\rm K}$ (as this brings the inner edge of the CM closer to the photosphere, where the magnetic field is stronger). Second, CB predicts a much higher plasma density for these stars, around $\log{N}\sim15-17$. This implies that some other physical mechanism must be operating to remove plasma from CMs. 

\bibliography{bib_dat}  % For BibTex

\end{document}